\documentclass[a4paper]{jpconf}

\usepackage{amssymb,amsfonts,amstext}

\newcommand{\lc}{\varepsilon}
\newcommand{\dlr}{\stackrel{\leftrightarrow}{d}{}}

\newcommand{\realni}{\ensuremath{\mathbb{R}}}
\newcommand{\celi}{\ensuremath{\mathbb{Z}}}
\newcommand{\ds}{\displaystyle}
\newcommand{\del}{\partial}
\newcommand{\goth}{\mathfrak}

\newcommand{\trr}{\triangleright}
\newcommand{\cf}{{\cal F}}
\newcommand{\cg}{{\cal G}}

\newcommand{\cL}{{\cal L}}
\newcommand{\cD}{{\cal D}}
\newcommand{\cM}{{\cal M}}

\begin{document}

\title{Categorical generalization of spinfoam models}
\author{Aleksandar Mikovi\'c$^{1,2}$ and Marko Vojinovi\'c$^{2}$}
\address{$^1$Departamento de Matem\'atica, Universidade Lus\'ofona de Humanidades e Tecnologias \\ $\,\,$ Av. do Campo Grande, 376, 1749-024 Lisboa, Portugal}
\address{$^2$Grupo de Fisica Matem\'atica da Universidade de Lisboa \\ $\,\,$ Av. Prof. Gama Pinto, 2, 1649-003 Lisboa, Portugal}

\ead{amikovic@ulusofona.pt, vmarko@cii.fc.ul.pt}

\begin{abstract}
We give a brief review of the problem of quantum gravity. After the discussion of the nonrenormalizability of general relativity, we briefly mention the main research directions which aim to resolve this problem. Our attention then focuses on the approach of Loop Quantum Gravity, specifically spinfoam models. These models have some issues concerning the semiclassical limit and coupling of matter fields. The recent developments in category theory provide us with the necessary formalism to introduce a new action for general relativity and perform covariant quantization so that the issues of spinfoam models are successfully resolved.
\end{abstract}

\section{Introduction}

It is well known that Einstein's theory of General Relativity is not straightforward to quantize. This is easily seen from the fact that GR is not perturbatively renormalizable. Simply put, one can attempt to quantize GR as an ordinary spin-two field in flat Minkowski spacetime, in the following way (for a nice review see \cite{Hamber}). Starting from the usual Einstein-Hilbert action 
$$
S_{EH} = \int d^4x\, \sqrt{-g} R,
$$
one rewrites the metric tensor $g_{\mu\nu}$ as the flat Minkowski metric $\eta_{\mu\nu}$ and the spin-two field $h_{\mu\nu}$ as
$$
g_{\mu\nu} = \eta_{\mu\nu} + h_{\mu\nu},
$$
and substitutes it into the action, rewriting it in terms of the new variable $h_{\mu\nu}$. Thereby one obtains
$$
\begin{array}{r}
\ds S_{EH} = \int d^4x\, h_{\mu\nu} \square h^{\mu\nu} + (\text{gauge fixing terms}) + \\
+ (\text{self-interaction terms}). \\
\end{array}
$$
The D'Alambertian operator is defined in the usual way, in flat Minkowski space, $\square \equiv \eta^{\mu\nu} \del_{\mu} \del_{\nu}$. From here one can proceed to perform the standard field theory quantization in the naive way --- first formulate the free quantum field theory, and then introduce interactions perturbatively.

However, very soon one is bound to face the difficulty of nonrenormalizability of this theory. The tree-level Feynman diagrams are finite, the one-loop divergences can be removed by wavefunction renormalization, but at the two-loop level a Lagrangian counterterm of the form
$$
\cL_2 = \frac{const}{\lc^2} R^{\alpha\beta}{}_{\mu\nu} R^{\mu\nu}{}_{\rho\sigma} R^{\rho\sigma}{}_{\alpha\beta} \qquad (\lc\to 0)
$$
appears \cite{GoroffSagnotti}, which is nonzero on-shell. Here $\lc = 4-D$ is the cutoff parameter from dimensional regularization scheme. At higher loop levels similar terms involving $R^4$, $R^5$, etc. terms are also expected to appear, rendering the theory perturbatively nonrenormalizable. This means that in order to remove all divergences one needs to introduce at least one additional coupling constant for each loop level. The infinite number of these coupling constants implies the loss of predictive power of the theory, since all experiments doable in principle can only ever fix a finite number of coupling constants.

This property of General Relativity has been known for quite some time, and there are various research directions which attempt to address this issue. They can be broadly separated into two classes, by the methodology.

The first class of approaches considers modifying or substituting GR by another theory, which should preferably be renormalizable. Such attempts have evolved into vast research directions such as supergravity, string field theory, noncommutative geometry, and so on. The goal of each proposed model is to have a renormalizable theory that looks like GR at least on the length scales which can be tested experimentally, while at the same time have only a finite number of coupling constants. These coupling constants could then in principle be used to predict the values of the infinite set of coupling constants appearing in the perturbative quantum gravity approach.

The second class of approaches is based on the point of view that abandons the renormalization paradigm, and essentially gives physical meaning to the cutoff parameters of some particular regularization scheme. In other words, the assumption is that at some scale (typically expected to be near the Planck scale) expectation values of the physical observables will start to depend explicitly on cutoff parameters. This dependence is assumed to be measurable (in principle), rather than being removed by renormalization. These attempts have also evolved into vast research directions such as loop quantum gravity, causal dynamical triangulations, causal set theory, etc. The goal of all proposed models is exactly the same as before --- predict some definite values for the infinite number of coupling constants present in the perturbative quantum gravity.

All these research directions have had limited success, and in the absence of any experimental data relevant at the Planck scale, none of these directions can be preferred over the others.

In what follows, we shall be mainly concerned with the approach of loop quantum gravity (for a review see \cite{Rovelli}), more specifically spin foam models, and we shall propose one novel particular model that addresses some serious issues present in all other spin foam models so far.

In section \ref{LQGSection} we shall give a short overview of the status of LQG in general and spin foam models in particular. We will argue that the main drawbacks of all 4D spin foam models stem from the fact that tetrad fields are not basic variables of the theory. Section \ref{TwoGroupSection} deals with the categorical generalization of the Poincar\' e group, called the Poincar\' e $2$-group. This will give us the necessary mathematical tools to reformulate the GR action in a convenient way which includes tetrad fields as basic variables. The analysis of this new action is then given in section \ref{SpincubeSection}, with a sketch of a quantization procedure giving rise to the so-called spincube model. Section \ref{DiscussionSection} contains conclusions and discussion of the results.

\section{\label{LQGSection}Loop Quantum Gravity and Spin Foam Mo\-dels}

A detailed review of the Loop Quantum Gravity approach can be found in \cite{Rovelli}. Here we just give some basic properties at an informal level.

The basic idea of LQG is to choose diffeomorphism-invariant quantities as basic degrees of freedom for the gravitational field, and then perform a canonical nonperturbative quantization of gravity in terms of these quantities. The natural candidates for basic variables turned out to be Wilson loops, and subsequently their generalizations called spin networks. This choice of variables introduces a natural diffeomorphism-invariant cutoff at the Planck length scale $l_P$, thereby rendering the theory UV-finite. The quantization is performed in the Schr\" odinger picture, and provides one with a mathematically well-defined constructions of the kinematical Hilbert space for the theory and some basic operators for geometric observables such as lengths, areas and volumes of space. Evolution in time is embodied in the Hamiltonian constraint, corresponding to the Wheeler-de Witt equation in the LQG setting.

The main features of such canonical approach to quantization are as follows. The theory represents a nonperturbative quantization of GR, and can in principle be applied to the study of physical systems where gravity is the dominant factor at short distances --- such systems include the black hole and cosmological singularities. It gives one a mathematical handle on a well-defined Hilbert space of states for the gravitational field, thereby giving some insight into the quantum mechanical features of gravity. The natural basis for the Hilbert space is the set of the {\em spin network states}, combinatorial graphs colored by the irreducible representations of the $SU(2)$ group, and corresponding intertwiners. Finally, the study of the geometric observables --- the length, area and volume operators --- reveals that each of them has a discrete spectrum, giving rise to the geometric interpretation of the gravitational field wavefunctional, as well as the discrete character of space.

The theory also has some drawbacks. First, the Hamiltonian constraint is not uniquely defined, due to the usual ordering problems present in quantum mechanics. Second, even if one chooses some particular ordering, the Hamiltonian constraint is extremely complicated and impossible to solve in practice. This severely limits the possibility for any practical calculations and the study of the dynamics of the theory. As the main obstacle, the proof of the correct semiclassical limit of the theory is still missing, as well as any attempt to predict the coupling constants from the perturbative gravity approach.

A way to resolve these drawbacks has been found in the spin foam approach \cite{RovelliSpinFoams}. The idea is to give up canonical quantization, but instead attempt a covariant, path-integral quantization of the theory. Building on the results of the canonical approach, one wants to define the gravitational path-integral
$$
Z = \int \cD g_{\mu\nu} \exp \left( i S_{EH}[g_{\mu\nu}] \right)
$$
in some way, in order to be able to calculate expectation values of observables, both in deep quantum regime and the semiclassical regime. This approach tends to give one a good handle on the dynamics of the theory, in addition to all features of the canonical approach.

The basic procedure of defining $Z$ goes as follows. One starts from the Plebanski action for General Relativity,
$$
S = \int B_{ab}\wedge R^{ab} + \phi^{abcd} B_{ab} \wedge B_{cd}.
$$
The first part of this action represents the topological $BF$ theory for the $SO(3,1)$ group. The $R^{ab}$ is the curvature $2$-form, a field strength ``$F$'' for the $SO(3,1)$ connection $1$-form $\omega^{ab}$. The $B_{ab}$ is the Lagrange multiplier $2$-form. The second part of the action is the Plebanski constraint, featuring $B_{ab}$ and the $0$-form Lagrange multiplier $\phi^{abcd}$. The purpose of the constraint is to enforce the $B_{ab}$ to be a simple $2$-form (i.e. an exterior product of two $1$-forms). This constraint is therefore called ``simplicity constraint'', and it can be shown that the simplicity requirement of the $B_{ab}$ field is enough to convert the topological $BF$ theory into General Relativity. The fact that $B_{ab}$ is simple gives rise to nontrivial degrees of freedom in the theory, reducing the equation of motion for $\omega^{ab}$ from Riemann-flat to Ricci-flat.

The second step is the quantization of the topological $BF$ theory. This can be done in a rigorous way by employing the methods of topological quantum field theory. One first discretizes spacetime into $4$-simplices, motivated by the structure of space in the canonical LQG, and rewrites the $BF$ action in the form
$$
\int B_{ab} \wedge R^{ab} \stackrel{\text{discr.}}{\longrightarrow} \sum_{\triangle} B_{\triangle} R_{\triangle},
$$
where the sum goes over all triangles in the triangulation. Then one defines a topological invariant
$$
\begin{array}{lcl}
Z & \equiv & \ds \int \cD\omega \int \cD B \exp \Big( i\sum_{\triangle} B_{\triangle} R_{\triangle} \Big) = \\
 & = & \ds \sum_{\Lambda} \prod_f A_2(\Lambda_f) \prod_v A_4(\Lambda_v). \\
\end{array}
$$
Here $\Lambda$ are the irreducible representations of $SO(3,1)$, labeling the faces $f$, edges $e$ and vertices $v$ of the Poincar\' e dual lattice corresponding to the triangulation. The colored $2$-complex dual to the spacetime triangulation is called a {\em spin foam}. The amplitudes $A_2(\Lambda)$ and $A_4(\Lambda)$ are determined such that $Z$ is in fact a topological invariant --- the total expression must not depend on the particular choice of the spacetime triangulation. In that way one arrives at the TQFT corresponding to the $BF$ theory for the $SO(3,1)$ group, commonly called the {\em Ooguri spin foam model}. Of course, the invariant $Z$ may be (and actually is) badly divergent, but that is not important at this stage, since we are only interested in the structure of the path integral.

The last step in the quantization procedure is to enforce the simplicity constraint on the $BF$ path integral at the quantum level. The exact technique for this is quite involved \cite{EPRL,FK}, but the bottom-line is that one projects the $SO(3,1)$ irreducible representations $\Lambda$ to the $SU(2)$ representations present in the canonical LQG formalism, in order to obtain the same structure of the Hilbert space on the spin foam boundary. The resulting theory is not topologically invariant, but represents one possible rigorous definition for the theory of quantum gravity. The most advanced spin foam model in this respect is the EPRL/FK model, developed independently by two research groups \cite{EPRL,FK}.

The main feature of spin foam models is that they correct some drawbacks of the canonical theory, primarily the dynamical sector is more under control. In addition, there remains a certain ambiguity in the choice of the amplitudes $A_2$ and $A_4$. This can be conveniently utilized to redefine the model such that it becomes IR-finite and to have a correct semiclassical limit \cite{MVefact,MVfiniteness}. One can also employ standard QFT methods to calculate the effective action for the model in the semiclassical limit, which opens a possibility to explicitly determine the coupling constants from perturbative quantum gravity.

Unfortunately, the spin foam models introduce their own set of problems. Aside from the ``unusual'' properties like fuzziness of geometry at the Planck scale, all spin foam models suffer from two major handicaps. The first is related to the fact that, in addition to the good semiclassical limit, all models have {\em additional semiclassical limits}, which do not give rise to the standard GR, but to the so-called {\em area-Regge geometry}. Since these different classical limits are not observed in experiments, one needs some additional mechanism to suppress such solutions. However, so far no mechanism could be constructed to deal with this problem.

The second handicap is related to the inability of the spin foam models to couple matter fields to gravity. Namely, the basic geometric variables which are employed in description of spacetime geometry are areas and volumes of space, but not lengths. This situation makes it extremely complicated (and in the case of massive fermionic matter even impossible) to incorporate matter fields into the spin foam model. Even if doable (see \cite{RovelliFermions} for the massless fermion coupling), the resulting theory would be too complicated to be useful for any calculation.

As it turns out, both of these handicaps have a common origin --- the edge lengths in the triangulation are not well-defined at the quantum level. This is itself a consequence of the choice of spin network states as basic degrees of freedom in the canonical LQG --- the choice which emphasizes the spin connection $\omega^{ab}$, while entirely ignoring the tetrad fields $e^a$. At the level of spin foam models, it is easy to see that the Plebanski constraint was purposefully designed to require the simplicity of $B_{ab}$, while avoiding to explicitly state that (the dual of) $B_{ab}$ is the product of two tetrad $1$-forms. The reason for this is that the tetrad fields do not appear as variables in the topological $BF$ sector of the theory, which is being used for the definition of the path integral.

In the remainder of this paper we will present a novel way to address this main difficulty, and to introduce tetrad fields explicitly in the topological sector of the theory. However, in order to do this, it is important to introduce some mathematical concepts which provide the background formalism for the new model.

\section{\label{TwoGroupSection}Poincar\' e 2-group}

We begin by giving a very brief review of the so-called {\em categorification ladder}, an important and active research topic in category theory. We shall not attempt at any rigor, leaving out most of the details, which can be found for example in \cite{BaezHuerta} and references therein.

In the branch of mathematics called {\em category theory}, one defines a structure called a {\em category} as a set of {\em objects} and a set of {\em morphisms} between those objects, satisfying some basic axioms. Such a structure is fairly general and does not have many interesting properties itself. However, this generality allows one to use it for all sorts of purposes. For example, one can define the usual structure of a {\em group} as a category which has only one object, while all morphisms (mapping the object onto itself) are invertible. The composition rules for the morphisms can be chosen to be the group multiplication, thereby providing an isomorphism between a given group and the corresponding category with one element.

The first step in the categorification ladder is to introduce the concept of a {\em 2-category}. A $2$-category consists of a set of objects, a set of morphisms and a set of {\em 2-morphisms}, maps between morphisms. Intuitively, if a category can be represented by a linear graph of dots (objects) and arrows connecting them (morphisms), a $2$-category can be represented by a planar graph, consisting of dots (objects), arrows connecting them (morphisms) and ``surface arrows'' mapping one arrow into another (see \cite{BaezHuerta} for details and pictures). The main point is that the dimensionality of the graph has been raised by one. The categorification ladder can continue by introducing a $3$-category (or in general an $n$-category) by a similar process, leading to $3$-dimensional (in general $n$-dimensional) graphs.

In analogy with a group, one can then define a {2-group}, as a $2$-category which has only one element, while all morphisms and $2$-morphisms are invertible. A $2$-group is a categorical generalization of a group, and is not a group itself. One can prove that any $2$-group is equivalent to a {\em crossed module}, a structure that has been studied independently by mathematicians before the idea of the categorification ladder has even been introduced. A crossed module is a quadruple $(G,H, \partial,\triangleright)$. This is a pair of groups $G$ and $H$, such that $\partial: H \to G$ is a homomorphism and $\triangleright : G\times H \to H$ is an action of $G$ on $H$ such that certain axioms are satisfied, which turn out to be directly related to the structure of a $2$-category, see \cite{BaezHuerta}. The elements of $G$ represent the $1$-morphisms, while the elements of the semidirect product $G\ltimes H$ represent the $2$-morphisms. The canonical example of a $2$-group relevant for physics is the Poincar\'e $2$-group, where $G = SO(3,1)$, $H={\realni}^4$, $\partial$ is a trivial homomorphism and $\triangleright$ is the usual action of the Lorentz transformations on the ${\realni}^4$ space. The Lorentz group is the group of morphisms, while the usual Poincar\'e group is the group of $2$-morphisms.

The main feature of the whole $2$-group formalism is that one can generalize the concept of a {\em holonomy} along a line to its two-dimensional analog --- a {\em surface holonomy}. The initial interest in this came from string theory. A point-particle travels along a world line in spacetime, and one is naturally led to the concept of a parallel transport along a given line. String theory promotes the point particle into a one-dimensional object --- a string --- which then travels along a world surface in spacetime. Thus one would like to have a concept of a {\em parallel transport along a given surface}. One of the main aims of the $2$-category and $2$-group formalism is to introduce and formalize this concept.

Given the strong categorical relationship between groups and $2$-groups, one can construct a gauge theory on a $4$-manifold $\cM$ based on a crossed module $(G,H, \partial,\triangleright)$ of Lie groups by using $1$-forms $A$, which take values in the Lie algebra $\goth g$ of $G$, and $2$-forms $\beta$, which take values in the Lie algebra $\goth h$ of $H$ \cite{GPP,fmm}. The forms $A$ and $\beta$ transform under the usual gauge transformations $g: \cM \to G$ as
$$
A \to g^{-1} A g \,+\,g^{-1} d g \,,\quad \beta\to g^{-1} \triangleright \beta \,,
$$
while the gauge transformations generated by $H$ are given by
$$
A \to A\,+\,\partial\eta\,,\quad\beta\to \beta +d \eta + A \wedge^\triangleright \eta + \eta\wedge \eta \, ,
$$
where $\eta$ is a one-form taking values in $\goth h$, see \cite{fmm}. When the group $H$ is Abelian, which happens in the Poincar\'e 2-group case, then the $\eta\wedge\eta$ term vanishes, and one obtains the gauge transformations given in \cite{GPP}.

The pair $(A,\beta)$ represents a $2$-connection on a $2$-fiber bundle associated to the $2$-Lie group $(G,H)$ and the manifold $\cM$. The corresponding curvature forms are given by
$$
{\cal F} = dA + A\wedge A -\partial\beta \,,\quad
{\cal G} = d \beta + A \wedge^\triangleright \beta \,,
$$
and they transform as
$$
{\cal F} \to g^{-1} {\cal F} g\,, \quad {\cal G} \to g^{-1} \trr {\cal G} \,,
$$
under the usual gauge transformations, while
$$
{\cal F} \to {\cal F}\,,\quad {\cal G}  \to {\cal G} +{\cal F} \wedge^\trr \eta\,,
$$
under the $H$-gauge transformations.

One can introduce a natural topological gauge theory determined by the vanishing of the 2-curvature
$$
\cf = 0 \,,\quad \cg =0 \,.
$$
These equations can be obtained from the action
$$
S =\int \langle B \wedge \cf \rangle_{\goth g} +  \langle C \wedge \cg \rangle_{\goth h} \,,
$$
where $B$ is a Lagrange multiplier $2$-form taking values in $\goth g$, $C$ is a Lagrange multiplier $1$-form taking values in $\goth h$, $\langle\;\;,\;\;\rangle_{\goth g}$ is a $G$-invariant nondegenerate bilinear form  in $\goth g$ and $\langle\;\;,\;\;\rangle_{\goth h}$ is a $G$-invariant nondegenerate bilinear form  in $\bf h$. This action is called $BFCG$ action, in analogy with the $BF$ theory action. The gauge transformations of the Lagrange multiplier fields are given by
$$
B \to g^{-1} B g \,,\quad C \mapsto g^{-1} \trr C \,,
$$
for the usual gauge transformations, while 
$$
B \to B - [C,\eta] \,,\quad C \mapsto C \,,
$$
for the $H$-gauge transformations.

Let us now examine the case of the Poincar\'e 2-group. In this case $A=\omega^{ab}J_{ab}$, $\beta = \beta^a P_a$, where $a,b\in\{0,1,2,3\}$,
$J_{ab}$ are the generators of the Lorentz group while $P_a$ are the generators of the translation group ${\realni}^4$. Consequently
$$
\begin{array}{cclcl}
\cf & = & \ds (d\omega^{ab} + \omega^a{}_c\wedge\omega^{cb} )J_{ab} & = & R^{ab}J_{ab} , \\
\cg & = & \ds \left(d\beta^a + \omega^a{}_b \wedge \beta^b \right) P_a & = & \left(\nabla\beta^a \right) P_a. \\
\end{array}
$$
The $G$-gauge transformations are the local Lorentz rotations
$$
\omega \to g^{-1} \omega g + g^{-1} dg \,,\quad \beta \to g^{-1} \trr \beta \,,
$$
while the $H$-gauge transformations are the local translations
$$
\delta_\varepsilon \omega^{ab} =0 \,,\quad \delta_\varepsilon \beta^a = d\varepsilon^a + \omega^a{}_b \wedge \varepsilon^b \,,
$$
where $\eta = \varepsilon^a P_a$.

The $BFCG$ action then becomes
$$
S = \int_{\cM} \left( B^{ab}\wedge R_{ab} + C_a \wedge \nabla\beta^a \right)\,, 
$$
where
$$
\delta_\varepsilon B = 0 \,,\quad \delta_\varepsilon C = 0 \,.
$$
At this point a very important observation is in order. The transformation properties of the $1$-form $C^a$ are the same as the transformation properties of the tetrad $1$-form $e^a$ under the local Lorentz and the diffeomorphism transformations. In addition, the equation of motion for $C^a$ is $\nabla C^a =0$, just like the no-torsion equation for the tetrad, $\nabla e^a =0$. Based on this, {\em we identify the Lagrange multiplier $C^a$ with the tetrad field $e^a$}, and write the action in the form
$$
S = \int_{\cM} \left( B^{ab}\wedge R_{ab} + e^a \wedge \nabla \beta_a \right) \,. \label{tga}
$$

In this way one can construct a categorical generalization of the topological $BF$ action. The new action is again topological, but more rich in structure, since the tetrad fields are explicitly present. In addition, the $2$-group formalism provides a framework to construct a topological quantum field theory from this action, in analogy with the $BF$ case. This provides us with the necessary tools to construct a categorical generalization of a spin foam model, based on the $BFCG$ action instead of the $BF$ action. The explicit presence of the tetrads should help us resolve the two handicaps of spin foam models discussed in section \ref{LQGSection}.

\section{\label{SpincubeSection}The Spincube Model}

The first step in the construction of the new model is to write the action for General Relativity, starting from the $BFCG$ action. In order to do this, all we need is the simplicity constraint,
$$
B_{ab} = \lc_{abcd}\, e^c\wedge e^d\,,
$$
which can now be added into the action as it stands, as opposed to the $BF$ case where the Plebanski constraint had to be introduced due to the absence of the tetrads $e^a$ in the $BF$ action. Therefore, one can write the {\em constrained $BFCG$ action} in the form
\begin{equation} \label{2pgr}
S = \int_{\cM} \Big[ B^{ab}\wedge R_{ab} + e^a \wedge \nabla \beta_a  -
\phi_{ab}\wedge \left( B^{ab} - \varepsilon^{abcd}e_c \wedge e_d \right) \Big]\,,
\end{equation}
where $\phi_{ab}$ is an additional Lagrange multiplier $2$-form field, introduced in order to enforce the simplicity constraint.

The equations of motion are obtained by varying $S$ with respect to $B$, $e$, $\omega$, $\beta$ and $\phi$, respectively, to give:
$$
\begin{array}{l}
R_{ab} - \phi_{ab} = 0\,,\\
\nabla\beta_a + 2\lc_{abcd} \phi^{bc}\wedge e^d = 0\,, \\
\nabla B_{ab} - e_{[a} \wedge \beta_{b]} = 0\,, \\
\nabla e_a = 0\,, \\
B_{ab} - \lc_{abcd} e^c \wedge e^d = 0\, .\\
\end{array}
$$
With the usual assumption that the tetrad fields are nondegenerate, these equations can be reworked into an equivalent form:
$$
\phi^{ab} = R^{ab}, \qquad B_{ab} = \lc_{abcd} e^c \wedge e^d, \qquad \beta^a = 0,
$$
$$
\nabla e^a = 0\,, \qquad \lc_{abcd} R^{bc}\wedge e^d = 0\,.
$$
The first three equations determine $\beta^a$ and the multipliers $B_{ab}$ and $\phi_{ab}$ in terms of $e^a$ and $\omega^{ab}$. The fourth equation is the no-torsion equation, which determines the connection $\omega^{ab}$ to be the Levi-Civita connection (a function of the tetrads $e^a$). The last equation is nothing but the Einstein field equation for the only remaining field $e^a$. Thus we see that the action (\ref{2pgr}) is classically equivalent to General Relativity. More precisely, it is equivalent to the Einstein-Cartan theory,
$$
S_{EC} = \int_{\cM} \varepsilon_{abcd} e^a \wedge e^b \wedge R^{cd} \,,
$$
since the torsion is equal to zero as an equation of motion rather than by definition.

Given the new action for General Relativity, we can proceed with the covariant quantization in analogy with the spin foam models. The action has the topological term and the constraint term, so as a first step we construct a topological quantum field theory by defining the path integral for the $BFCG$ part of the action. In the second step, we enforce the constraint term by requiring a suitable restriction in the path integral of the topological theory.

One begins by triangulating spacetime into $4$-simplices, and rewriting the topological part of the action in the form
$$
\sum_{\triangle} B_{\triangle} R_{\triangle} + \sum_{l} e_l ( \nabla\beta )_l,
$$
where the first sum goes over all triangles and the second goes over all edges in the triangulation of the spacetime manifold. Then one constructs a topologically invariant path integral in the form (see \cite{MVtwoPoincare} for the details of the construction)
\begin{equation} \label{StateSuma}
\begin{array}{lcl}
Z & \equiv & \ds \int \cD\omega \int \cD B \int \cD e \int \cD \beta \,
 \exp \Big( i\sum_{\triangle} B_{\triangle} R_{\triangle} + i \sum_{l} e_l ( \nabla\beta )_l \Big) = \\
 \vphantom{\ds\int} & = & \ds \sum_{\Lambda} \prod_p A_1(\Lambda_p) \prod_f A_2(\Lambda_f) \prod_v A_4(\Lambda_v). \\
\end{array}
\end{equation}
The labels $\Lambda = (L_p,m_f)$, where $L_p \in\realni_0^+$ and $m_f\in\celi$, are now irreducible representations of the Poincar\' e $2$-group, and in addition to vertices $v$ and faces $f$ of the Poincar\' e dual lattice, we also take the product over all the polyhedra $p$, since they are dual to the edges of the triangulation and naturally appear in the construction due to the presence of the $e\wedge \nabla\beta$ term in the $BFCG$ action. The amplitudes $A_1(\Lambda)$, $A_2(\Lambda)$ and $A_4(\Lambda)$ are chosen so that $Z$ does not change under the action of the Pachner moves, which guarantees its independence of the triangulation. The polyhedra are colored with $L_p$, which have the interpretation as lengths of triangulation edges, while faces are colored with $m_f$, which have the interpretation as areas of the triangles in the triangulation. In the topological theory, edge lengths and triangle areas are independent of each other.

Note that the path integral is not defined over a colored $2$-complex (the spinfoam), but rather over a colored $3$-complex (called {\em spincube}).

Finally, we can impose the simplicity constraint, in order to turn the topological path integral into a realistic model for quantum gravity. Based on the geometric interpretation of the variables, the constraint actually says that a very natural requirement should be enforced --- the triangle areas must be compatible with the corresponding edge lengths. This can be formalized in the requirement
$$
|m_f|l_P^2 = A_f (L) , \quad \forall f
$$
where $A_f(L)$ is the Heron formula for the triangle area in terms of its edges. The Planck length appears naturally in order to balance the dimensions of the two sides of the equation. As a last step, one redefines the amplitudes $A_1$, $A_2$ and $A_4$ in order to render the theory IR-finite, as well as to enforce the correct semiclassical limit, in a way similar to the spinfoam models.

Note that imposing this constraint leaves only edge lengths as independent variables in the theory, so that the ``area-Regge'' problem present in spinfoam models is resolved automatically. In addition, the edge length variables allow for a completely straightforward coupling of matter fields to the spincube model. Namely, at the level of the classical theory, one can introduce fermions via the action
\begin{equation} \label{DejstvoSaFermionom}
\begin{array}{ccl}
S & = & \ds \int \Big[ B^{ab}\wedge R_{ab} + e^a \wedge \nabla \beta_a  - \phi_{ab}\wedge \left( B^{ab} - \varepsilon^{abcd}e_c \wedge e_d \right) \Big] + \\
 & & \ds \hphantom{mm} + i \kappa_1 \int \lc_{abcd} \, e^a \wedge e^b \wedge e^c \wedge \bar{\psi} \left[ \gamma^d  \dlr + \{ \omega , \gamma^d \}  + \frac{im}{2}\,e^d \right] \psi + \\
 & & \ds  \hphantom{mmmmmmm} + i\kappa_2 \int \lc_{abcd} e^a \wedge e^b \wedge \beta^c \, \bar{\psi}\gamma_5\gamma^d \psi \, , \\
\end{array}
\end{equation}
where $\omega = \omega_{ab} [\gamma^a , \gamma^b]/8$, $\kappa_1 = 8\pi l_P^2 /3$ and $\kappa_2 = - 2 \pi l_P^2 $. The first term is the constrained $BFCG$ action, while the second and third terms introduce fermion coupling which results in the same equations of motion as in the ordinary Einstein-Cartan theory with fermions.

The quantization procedure of the action (\ref{DejstvoSaFermionom}) is essentially the same as the one without fermions. The only difference is in the fact that the vertex amplitude $A_4$ will change to reflect the presence of the fermionic matter, as
$$
A_4 \to A_4 \exp\left[ iS_R^{(\rm ferm)}(L,\psi) \right] \, ,
$$
where $S_R^{(\rm ferm)}$ is the Regge discretized action of a fermion field $\psi$ coupled to gravity. The expressions which appear in $S_R^{(\rm ferm)}$ can be easily obtained, in contrast to the EPRL/FK model case, where the expression for the $4$-simplex volume is impossible to define uniquely in terms of the spin foam variables \cite{RovelliFermions}.

Similarly to (\ref{DejstvoSaFermionom}), one can also couple other matter fields to (\ref{2pgr}) in a completely straightforward way, including gauge and scalar fields, the cosmological constant, the Holst term, and so on.

\section{\label{DiscussionSection}Conclusions}

The proposed 2-group reformulation of GR can be used to obtain a categorical ladder generalization of Loop Quantum Gravity. The advantage of this generalization is that the edge lengths of a triangulation become the basic dynamical variables. This will facilitate the construction of the path integral such that the classical limit of the corresponding quantum theory is GR and the coupling of matter will be much easier to accomplish. 

The categorical nature of the theory implies that the edge labels of a spacetime triangulation should be the 2-group irreducible representations on a $2$-Hilbert space. Note that this is not unique, since one can also use the category of chain complexes of vector spaces in order to define the representations, see \cite{fmm,cfm}. The structure of the chain-complex representations is different from the 2-Hilbert space representations, which means that chain-complex representation theory defines an alternative quantization of GR. Hence it would be interesting to develop the chain-complex representation theory of the Poincar\' e 2-group.

The physical significance of 2-Hilbert space representations could be better understood by performing a canonical quantization of the action (\ref{2pgr}).

As far as the construction of 4-manifold invariants based on the $BFCG$ state sum is concerned, one would have to regularize the topological state sum/integral based on the amplitude (\ref{StateSuma}) such that the triangulation independence is preserved. One way to do it is to try to implement a gauge-fixing procedure, see \cite{bafr}. Another way is to find a quantum group regularization, since there are strong indications that categorified quantum groups and their representations will be important for the construction of $4$-manifold invariants \cite{crfr}. Hence one can try to find a crossed module of Hopf algebras which is a deformation of the Poincar\' e 2-group, and then try to find an appropriate $2$-category of representations which will give a finite topological state sum.

\ack

This work has been partially supported by FCT project PTDC/MAT/099880/2008. MV was also supported by the FCT grant SFRH/BPD/46376/2008.

\end{document}